\documentclass[conference]{IEEEtran}
	\usepackage{amsmath,amsfonts}
	\usepackage{bm}
	\usepackage{algorithmic}
	\usepackage{hyperref}
	\usepackage{algorithm}
	\usepackage[caption=false,font=footnotesize,labelfont=rm,textfont=rm]{subfig}
	\usepackage{textcomp}
	\usepackage{stfloats}
	\usepackage{url}
	\usepackage{verbatim}
	\usepackage{epsfig}
	\usepackage{color}
	\usepackage{soul}
	\usepackage{multirow,makecell}
	\usepackage{amssymb}
	\usepackage{graphicx}
\usepackage{algorithm}
	\def\BibTeX{{\rm B\kern-.05em{\sc i\kern-.025em b}\kern-.08em
			T\kern-.1667em\lower.7ex\hbox{E}\kern-.125emX}}
\begin{document}

\title{{Secure Transmission }for Movable Antennas Empowered Cell-Free Symbiotic Radio Communications}
		

\author{
	\IEEEauthorblockN{
	Jiayu Guan\textsuperscript{1}, Bin Lyu\textsuperscript{1}, Yan Liu\textsuperscript{2}, and Feng Tian\textsuperscript{1}}
\IEEEauthorblockA{\textsuperscript{1}
Nanjing University of Posts and Telecommunications, Nanjing 210003, China}

\IEEEauthorblockA{\textsuperscript{2}  Tongji University, Shanghai 201804, China}
}
		
\markboth{IEEE Communications Letters}%
{Shell \MakeLowercase{\textit{et al.}}: A Sample Article Using IEEEtran.cls for IEEE Journals}

\maketitle
		
\begin{abstract}
In this paper, a novel movable antenna (MA) empowered secure transmission scheme is designed for cell-free symbiotic radio (SR) systems in the presence of an eavesdropper (Eve). Specifically, multiple distributed access points (APs) equipped with MAs collaboratively transmit confidential information to the primary user (PU), {in the meanwhile the backscatter device (BD) transmits its} own information to the secondary user (SU) by reflecting incident signals from the APs. The MAs deployed at the APs can adjust their positions flexibly to improve channel conditions between the APs and the PU/SU/BD and suppress the eavesdropping from the Eve on confidential information at the PU. Under this setup, {we maximize} the secrecy rate of primary  transmission through {jointly optimizing }the APs' transmission beamforming vectors and the positions of the MAs, while adhering to {the quality of service constraints at the SU. To address {the challenges} caused by the non-convexity and find a near-optimal solution,} an alternating optimization (AO) framework is proposed, utilizing the successive convex approximation method, the semi-definite relaxation technology and a genetic algorithm modified particle swarm optimization (GA-PSO) algorithm. Numerical results demonstrate {the secrecy rate enhancement} provided by utilizing the MAs and show the impact of the GA-PSO algorithm for improving the solving accuracy.

\end{abstract}
		

\section{Introduction}
		
\IEEEPARstart{A}{s} wireless networks advance, the number of Internet of Things (IoT) devices is expected to {rise dramatically.} Consequently, there will be a substantial demand for energy and spectrum resources to facilitate their interconnection \cite{YC}. However, the insufficient availability of energy and spectrum resources {restricts the deployment of} IoT devices. {To resolve the aforementioned limitation,} symbiotic radio (SR) has been proposed and already garnered significant interest in the academic community \cite{LYC}.

In typical SR communication systems, the secondary  transmission system shares spectrum and power with the primary  transmission system through reflection technology, thereby improving {energy and spectral efficiency} \cite{LYC}. In \cite{LYC}, SR was systematically introduced to emphasize its enormous potential in future wireless communication networks. In \cite{LongIoT}, the achievable rate performance was been analyzed under the parasitic SR (PSR) setup, in which both primary and secondary signals share identical durations. {Besides, the authors in} \cite{HY} {focused on maximizing system energy efficiency (EE) for the SR system.}
However, the above works focused on traditional cellular architectures, where inter-cell interference {is recognized as a constraint on supporting stable high-quality services} \cite{DDZZ}. To address this constraint, cell-free networking has been seen as a promising architecture. Motivated by this, the cell-free SR system was investigated in \cite{DDZZ}, {where an efficient channel estimation method is proposed based on two-phase uplink training.} {In \cite{FL}, the spectral efficiency (SE) of uplink cell-free massive MIMO SR systems was investigated.}

However, {the open nature of wireless networks introduces 
secure transmission issues.} For example, the confidential information transmitted by the transmitter can be intercepted by eavesdroppers (Eves).
Therefore, physical layer security (PLS) mechanisms \cite{PLS1} have been developed to address this concern. {The authors of} \cite{ST} {proposed a transmit power allocation algorithm aimed at maximizing the achievable secrecy rate in a cell-free MIMO system.} In \cite{AAI}, a transmission scheme incorporating artificial noise (AN) has been studied {to ensure the secure transmissions in SR systems. However, based on our current knowledge,} the secure transmission issues have been overlooked in cell-free SR communication systems.

Moreover, in \cite{LongIoT}-\cite{AAI}, the transmitters were equipped with fixed position antennas (FPAs), which {leads to} a dilemma where spatial degrees of freedom (DoFs) cannot be fully leveraged \cite{ma2023}. More importantly, the steering vector of the FPAs remains static, which may weaken the security beamforming gain. To handle these challenges, the concept of movable antenna (MA) has been proposed \cite{ma2023}. Unlike reconfigurable intelligent surface configuring channel conditions by adjusting the amplitude and phase shift of incident signals \cite{QQQ},  the MA technology achieves the same goal by moving antennas to appropriate positions. The real-time positioning adjustment of MAs is enabled by their connection to radio frequency (RF) chains via flexible cables \cite{ma2023}.  In \cite{LZHUMA}, the channel model of MA-assisted systems was studied. In \cite{PLS}, the MA technology was demonstrated to be an efficient method to achieve  secure wireless communications.
 In \cite{LYUB111} and \cite{ZC}, {the MA technology was leveraged in SR systems for improving the performance both primary and secondary communication systems. However, how to apply MAs in cell-free SR systems for simultaneously enhancing system performance and achieving secure transmission is still unknown.}

In this paper, we investigate an MA empowered {PLS mechanism} for the cell-free SR system in the presence of an Eve. {In the system,} multiple distributed APs equipped with MAs collaboratively send confidential information to the  primary user (PU) to resist eavesdropping from the Eve. {At the same time,} the {backscatter device (BD)} achieves secondary transmission by reflecting incident signals from APs to transmit its own information to the secondary user (SU). {The MAs applied at the APs can flexibly adjust their positions to improve the channel conditions {associated with} the primary and secondary communications and worsen the transmission links to the Eve. Under this setup,} we consider the problem of maximizing the secrecy rate of primary transmission for the PU under {the quality of service (QoS) constraints imposed on the SU.} To solve this non-convex optimization problem, {an alternating optimization (AO) framework is adopted to decompose the coupling between the transmit beamforming vectors and the position variables of the MAs. } For designing the transmit beamforming, we employ the successive convex approximation (SCA) and the semidefinite relaxation (SDR) {to derive a near-optimal solution.} For optimizing the positions of MAs, we utilize a genetic algorithm modified particle swarm optimization (GA-PSO) algorithm. {Numerical results showcase the superior performance of the proposed MA empowered scheme with the GA-PSO algorithm in enhancing the secrecy rate.}

%
%

\section{System Model}
We consider an MA empowered cell-free SR communication system, which consists of $M$ distributed APs, one PU, one SU, and one BD. Each AP is equipped with $N$ MAs, the BD has $L$ FPAs, and the other devices are each with single FPA. In the considered system, there also exists an Eve with single FPA around the PU for intercepting the information transmitted from the APs to the PU. However, due to the obstacles between the SU and the Eve, the Eve cannot intercept the information at the SU \cite{ZHOUC}. {The system model is illustrated in} Fig. \ref{fig:system_model}. By utilizing flexible cables to connect the RF chains and MAs at each AP, the MAs can adaptively move within a certain range to construct favorable {transmission links} \cite{LYUB111}. {The sets of the MAs at each AP, the APs and the FPAs at the BD are defined as} $\mathcal{N}= \{1\text{,}\ldots\text{,} N\}$, $\mathcal{M}= \{1\text{,}\ldots\text{,} M\}$, and $\mathcal{L}= \{1\text{,}\ldots\text{,} L\}$, respectively. {A three-dimensional coordinate system is leveraged to model the positions.} Specifically, {we define} $\bm{t}_{m\text{,}a} =[x_{m}\text{,} y_{m}\text{,} z_{m}]^T$ as the $m$-th AP's position, where $m \in \mathcal{M}$. The position of all $N$ MAs at the $m$-th AP is represented by $\bm{t}_m =[\bm{t}_{m\text{,}1}^T\text{,} \cdots \text{,} \bm{t}_{m\text{,}N}^T]^T$, where $\bm{t}_{m\text{,}n} =[x_{m\text{,}n}\text{,} y_{m\text{,}n}\text{,} z_{m\text{,}n}]^T$ is the position of the $n$-th MA at the $m$-th AP, and $n \in \mathcal{N}$. Let $\bm{t} =[\bm{t}_1^T\text{,} \cdots \text{,} \bm{t}_M^T]^T$ represent all MAs' positions. 
The position of the $L$ FPAs at the BD can be denoted by $\bm{r}_{b} =[\bm{r}_{l}^T\text{,} \cdots\text{,} \bm{r}_{L}^T]^T$, where $\bm{r}_{l} =[x_{l}\text{,} y_{l}\text{,} z_{l}]^T$ is the position of the $l$-th FPA {{at the BD, and  $l \in \mathcal{L}$. Similarly, the positions of the FPAs} at the PU, SU and Eve {are fixed and denoted by} $\bm{r}_{p} =[x_{p}\text{,} y_{p}\text{,} z_{p}]^T$, $\bm{r}_{s} =[x_{s}\text{,} y_{s}\text{,}z_{s}]^T$, and $\bm{r}_{e} =[x_{e}\text{,} y_{e}\text{,} z_{e}]^T$, respectively. 
\begin{figure}
  \centering
  \subfloat[]{
  	\includegraphics[scale=0.68]{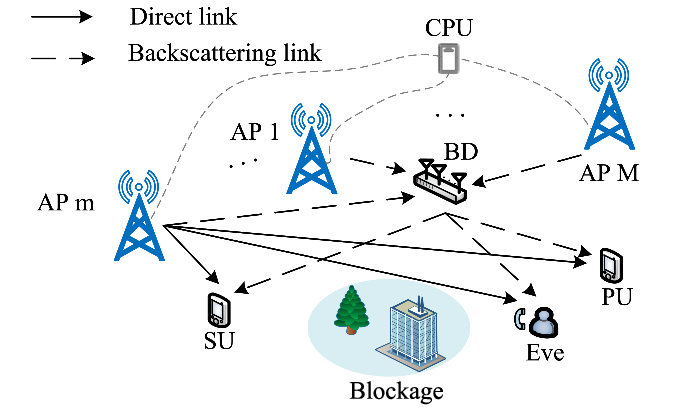}}\\
  \subfloat[]{
  	\includegraphics[scale=0.68]{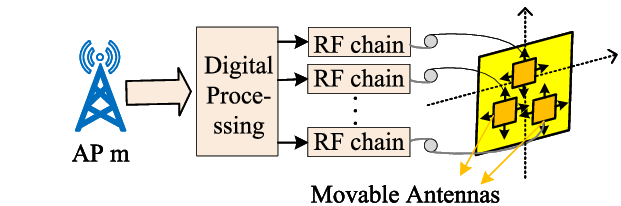}}
  \caption{The MA empowered cell-free SR communication system. (a): System model. (b): Illustration of MAs at the $m$-th AP. }
  \label{fig:system_model}
\end{figure}

\subsection{Channel model}
{Assume that} the channels remain static in {the considered transmission frame} for given positions of MAs.
{Consider $L_{t\text{,}\kappa}^m$ and $L_{r\text{,}\kappa}^m$ as the number of transmit and receive channel paths
from the $m$-th AP to the node-$\kappa$, respectively, 
where $\kappa = \{l\text{,}p\text{,}e\text{,}s\}$ {indicates} that the node is the BD/PU/Eve/SU. Similarly, $L_{t\text{,}\xi}^b$ and $L_{r\text{,}\xi}^b$ denote the quantity of transmit and receive channel paths between the BD and the node-$\xi$, respectively, where $\xi = \{p\text{,}e\text{,}s\}$ represents that the node is the PU/Eve/SU.}
Denote the sets of transmit and receive channel paths
from the $m$-th AP to the node-$\kappa$ as $\mathcal{L}_{t\text{,}\kappa}^m=\{1\text{,}\cdots\text{,}L_{t\text{,}\kappa}^m\}$ and $\mathcal{L}_{r\text{,}\kappa}^m=\{1\text{,}\cdots\text{,}L_{r\text{,}\kappa}^m\}$,
respectively. 
Let $\mathcal{L}_{t\text{,}\xi}^b=\{1\text{,}\cdots\text{,} L_{t\text{,}\xi}^b\}$ and $\mathcal{L}_{r\text{,}\xi}^b=\{1\text{,}\cdots\text{,}L_{r\text{,}\xi}^b\}$
denote the sets of transmit and receive channel paths between the BD and the node-$\xi$, respectively.

Following \cite{ma2023}, {we adopt the far-field wireless channel model because of the substantial size difference between the transmit/receive region and the signal propagation distance. Consequently, the angle of departure (AoD) and angle of arrival (AoA) {are same} for all MAs within the transmit/receive region} \cite{ma2023}.
The azimuth and elevation AoDs of the $j_1$-th transmit path between the $m$-AP and the node-$\kappa$ are denoted by $\phi_{m\text{,}j_1}^t$ and $\theta_{m\text{,}j_1}^t$, respectively, where $j_1\in \mathcal{L}_{t\text{,}\kappa}^m$. 
 The signal propagation difference between the $\bm{t}_{m,n}$ and the origin of the transmit region for the $j_1$-th transmit path is expressed as {
$\rho_{m\text{,}j_1}^t(\bm{t}_{m\text{,}n}) = x_{m\text{,}n} \cos\theta_{m\text{,}j_1}^t \cos\phi_{m\text{,}j_1}^t + y_{m\text{,}n}\cos\theta_{m\text{,}j_1}^t \sin\phi_{m\text{,}j_1}^t + z_{m\text{,}n}\sin\theta_{m\text{,}j_1}^t$} \cite{ZHUma}.
Similarly, 
the signal propagation difference between position $\bm{r}_{\kappa}$ and the origin of the receive region of the $i_1$-th receive path is denoted by {
$\rho_{\kappa\text{,}i_1}^r(\bm{r}_\kappa) = x_\kappa \cos\theta_{\kappa\text{,}i_1}^r \cos\phi_{\kappa\text{,}i_1}^r + y_\kappa \cos\theta_{\kappa\text{,}i_1}^r \sin\phi_{\kappa\text{,}i_1}^r + z_\kappa \sin\theta_{\kappa\text{,}i_1}^r$,}
where $\phi_{\kappa\text{,}i_1}^r$  and $\theta_{\kappa\text{,}i_1}^r$ are the azimuth and elevation AoAs, and $i_1\in \mathcal{L}_{r\text{,}\kappa}^m$.

Based on the descriptions above, we first model the channel between the $m$-AP and the BD, {denoted by} $\bm{H}_{m}^H $, as
\begin{align}
	\bm{H}_{m}^H = \bm{F}_{m\text{,}l}^H \bm{\Sigma}_l \bm{K}_{m\text{,}l}\text{,}
\end{align}
where $\bm{\Sigma}_l \in \mathbb{C}^{L_{r\text{,}l}^m \times L_{t\text{,}l}^m}$ is the path-response matrix, $\bm{F}_{m\text{,}l}(\bm{r}_b) = [\bm{f}_{m\text{,}l}(\bm{r}_{1})\text{,} \cdots\text{,} \bm{f}_{m\text{,}l}(\bm{r}_{L})]\in \mathbb{C}^{L_{r\text{,}l}^m \times L}$ is the receiving field response matrix, $\bm{f}_{m\text{,}l}(\bm{r}_{l}) = [e^{j \frac{2\pi}{\lambda}\rho_{l\text{,}1}^r(\bm{r}_{l})}\text{,}\cdots\text{,} e^{j \frac{2\pi}{\lambda}\rho_{l\text{,}L_{r\text{,}l}^m}^r(\bm{r}_{l})} ]^T \in\mathbb{C}^{L_{r\text{,}l}^m \times 1}$ is the receiving field response vector (FRV),  $\lambda$ is the carrier wavelength, $\bm{K}_{m\text{,}l}(\bm{t}_m) = [\bm{k}_{m\text{,}l}(\bm{t}_{m\text{,}1})\text{,} \cdots\text{,} \bm{k}_{m\text{,}l}(\bm{t}_{m\text{,}N})]\in \mathbb{C}^{L_{t\text{,}l}^m \times N}$ is the transmit field-response matrix, and $\bm{k}_{m\text{,}l}(\bm{t}_{m\text{,}n}) = [e^{j \frac{2\pi}{\lambda}\rho_{m\text{,}1}^t(\bm{t}_{m\text{,}n})}\text{,}  \cdots\text{,}e^{j \frac{2\pi}{\lambda}\rho_{m\text{,}L_{t\text{,}l}^m}^t(\bm{t}_{m\text{,}n})} ]^T \in\mathbb{C}^{L_{t\text{,}l}^m \times 1}$ is the transmit FRV.

Then, the channel from the $m$-AP to the node-$\xi$ can be modeled as
\begin{align}
\bm{h}_{m\text{,}\xi}^H = \bm{f}_{m\text{,}\xi}^H \bm{\Sigma}_\xi \bm{K}_{m\text{,}\xi}\text{,}
\end{align}
where $\bm{\Sigma}_\xi \in \mathbb{C}^{L_{r\text{,}\xi}^m \times L_{t\text{,}\xi}^m}$ is the path-response matrix, $\bm{f}_{m\text{,}\xi}(\bm{r}_\xi) = [e^{j \frac{2\pi}{\lambda}\rho_{\xi\text{,}1}^r(\bm{r}_\xi)}\text{,} \cdots\text{,} e^{j \frac{2\pi}{\lambda}\rho_{\xi\text{,}L_{r\text{,}\xi}^m}^r(\bm{r}_\xi)} ]^T \in\mathbb{C}^{L_{r\text{,}\xi}^m \times 1}$ is the receiving FRV,  $\bm{K}_{m\text{,}\xi}(\bm{t}_m) = [\bm{k}_{m,\xi}(\bm{t}_{m\text{,}1})\text{,} \cdots\text{,} \bm{k}_{m\text{,}\xi}(\bm{t}_{m\text{,}N})]\in \mathbb{C}^{L_{t\text{,}\xi}^m \times N}$ is the transmit field-response matrix, and $\bm{k}_{m\text{,}\xi}(\bm{t}_{m\text{,}n}) = [e^{j \frac{2\pi}{\lambda}\rho_{m\text{,}1}^t(\bm{t}_{m\text{,}n})}\text{,} \cdots\text{,} e^{j \frac{2\pi}{\lambda}\rho_{m\text{,}L_{t\text{,}\xi}^m}^t(\bm{t}_{m\text{,}n})} ]^T \in\mathbb{C}^{L_{t\text{,}\xi}^m \times 1}$ is the transmit FRV.

Finally, we can model the channel from the BD to the node-$\xi$ as
$\bm{g}_{b,\xi}^H \in \mathbb{C}^{1 \times L}$ {since the positions of antennas at the BD and the node-$\xi$ are fixed.}

\subsection{Transmission model}
The PSR setup is considered, based on {which the primary and secondary symbols have equal durations} \cite{LongIoT}. The primary and secondary symbols  are represented by $s(n) \sim \mathcal{CN}(0\text{,}1)$ and $c(n) \sim \mathcal{CN}(0\text{,}1)$, respectively. Let $\bm{w}_m \in \mathbb{C}^{N \times 1}$ denote the transmit beamforming vector of the $m$-th AP, {under the constraint that} $||\bm{w}_m||^2 \le P_\text{max}$, and {the maximum transmit power offered by each AP is denoted by $P_\text{max}$.} {We can express the received signal at the node-$\xi$ as} \cite{DDZZ} 
 \begin{align}
 	\label{y}
 	y_{\xi}(n) &= \sum_{m=1}^M\bm{h}_{m,\xi}^H \bm{w}_ms(n) \notag\\&+ \sum_{m=1}^M\sqrt{\alpha} \bm{g}_{b,\xi}^H \bm{H}_{m}^H \bm{w}_m s(n) c(n)+z_\xi(n)\text{,}
 \end{align} 
where $\alpha \in [0\text{,}1]$ is the reflection coefficient of the BD, and  $z_\xi(n) \sim \mathcal{CN}(0\text{,}\sigma_\xi^2)$ is the additive Gaussian white noise at the node-$\xi$. 

When the PU/Eve decodes the primary symbol $s(n)$, the term involved with $c(n)$ {is regarded as interference, since the signal strength of the backscattering link is significantly weaker compared to the direct link} \cite{DDZZ}. Then, the achievable rate of primary communication from the APs to the node-$\xi$ is 
\begin{align}
	\label{R}
R_{\xi} = B\log_2(1+\gamma_{\xi})\text{,} ~\xi \in \{p\text{,}e\text{,}s \}\text{,}
\end{align}
where $B$ represents the channel bandwidth, 
$\gamma_{\xi} = \frac{|\sum_{m=1}^M\bm{h}_{m,\xi}^H \bm{w}_m|^2 }{\alpha|\sum_{m=1}^M\bm{g}_{b,\xi}^H \bm{H}_{m}^H \bm{w}_m|^2 + \sigma_\xi^2 }$ 
denotes the signal to interference plus noise ratio (SINR).
The secrecy rate of primary communication for the PU is denoted by
\begin{align}
  R_\text{sec}=[R_p-R_e]^+ \text{,}
\end{align}
where $[\hat{b}]^+=\text{max}\left\{ \hat{b}\text{,}0 \right\}$.

 {The SU also {needs to} decode both the primary and secondary signals.} The first term in \eqref{y} {can be} subtracted after that the primary signal is decoded as shown in \eqref{R} \cite{DDZZ}. The resulting signal is expressed as 
\begin{align}
 	y_{c}(n) = \sqrt{\alpha}\sum_{m=1}^M[  \bm{g}_{b\text{,}s}^H \bm{H}_{m}^H \bm{w}_m s(n) c(n)]+z_s(n).
\end{align}
 The primary signal $s(n)$ serves as the fast-varying channel response, when decoding the secondary signal $c(n)$ \cite{LongIoT}. {Thus, the instantaneous} channel gain of the fast-varying channel is primarily dominated by $|s(n)|^2$. Therefore, the ergodic rate of the link from the BD to the SU can be formulated as \cite{DDZZ}
\begin{align}
 R_c &= B\mathbb{E}_{s(n)}[\log_2(1+\beta_c|s(n)|^2) ]
 \notag\\&=\int_{0}^{\infty}\log_2(1+\beta_cx)e^{-x}dx\notag\\&=-e^{\frac{1}{\beta_c}}\text{Ei}(-\frac{1}{\beta_c})\log_2e\text{,}
\end{align}
where $\text{Ei}(x) = \int_{-\infty}^{x}(\frac{1}{u})e^udu$ denotes the exponential integral, and $\beta_c = \frac{\alpha|\sum_{m=1}^M\bm{g}_{b\text{,}s}^H\bm{H}_{m}^H \bm{w}_m|2}{\sigma_s^2}$ is the corresponding signal-to-noise ratio (SNR).

\section{Problem formulation and proposed solution}
The aim of this section is to maximize {the secrecy rate at the PU} {through joint optimization of} the transmit beamforming vectors at the APs and the positions of the MAs, under constraints $\text{C1:~} R_s \geq R_\text{th1}$, $\text{C2:~} R_c \geq R_\text{th2}$, $\text{C3:~}  ||\bm{w}_m ||_2^2 \le P_{\text{max}}$, $\text{C4:~}  \left\|\bm{t}_{m\text{,}i} - \bm{t}_{m\text{,}j}\right \|_2\geq D$, and $\text{C5:~}  x_{m\text{,}n}^{\text{min}} \leq x_{m\text{,}n} \leq x_{m\text{,}n}^{\text{max}}\text{,}\ y_{m\text{,}n}^{\text{min}} \leq y_{m\text{,}n} \leq y_{m\text{,}n}^{\text{max}}\text{,}\ z_{m\text{,}n}^{\text{min}} \leq z_{m\text{,}n} \leq z_{m\text{,}n}^{\text{max}}$. $\text{C1}$ and $\text{C2}$ indicate that the achievable rates of decoding the primary and secondary signals at the SU should satisfy the given threshold, i.e., $R_\text{th1}$ and $R_\text{th2}$. {C3 is the power constraint applied to the $m$-th AP.} $\text{C4}$ is the minimum inter-MA distance constraint to avoid coupling effects between each pair of MAs. In addition, $\text{C5}$ restrains the movable region of the MAs. {The formulation of the optimization problem is as follows,}
\begin{equation}\tag{$\textbf{P1}$} 
		\begin{aligned}
			   	\max_{\bm{w}_m\text{,} \bm{t}_m} ~~&  R_\text{sec} \\ 
				\text{s.t.}~~ & \text{C1}\text{,}\ \text{C2}\text{,}\ \text{C3}\text{,}\ \text{C4}\text{,}\ \text{C5}.
			\end{aligned}
\end{equation}

 {Obviously}, \textbf{P1} suffers from the non-convexity {due to the non-convex nature of both the objective function and constraints $\text{C1}$-$\text{C4}$.} We thus propose an AO framework to address this challenge, in which the SCA and the SDP methods are explored to optimize the APs' transmit beamforming vectors, and {the GA-PSO algorithm is leveraged to update the MAs' positions} \cite{GAPSO1}.

\subsection{Optimization of the transmit beamforming vector}
	\label{OpTB}
In this subsection, we start by fixing the positions of the MAs} and then optimize the APs' transmit beamforming vectors to maximize $R_\text{sec}$. As shown in \cite{DDZZ}, {as $\beta_c$ increases, $R_c$ is monotonically non-decreasing.
Thus, the optimal solution of $R_c= R_\text{th2}$, {denoted by $\beta_c^*$, can be derived} by utilizing the bisection method. Based on this, constraint $\text{C2}$ is rewritten as $\text{C6:~}\beta_c\geq \beta_c^*$. Furthermore, we introduce a semi-definite matrix $\bm{W}=\bm{w}\bm{w}^H$ with $\text{C10:~}\text{Rank}(\bm{W})=1$ and $\text{C11:~}\bm{W} \succeq 0$, where  $\bm{w}^T=[\bm{w}_1^T\text{,}\bm{w}_2^T\text{,}\cdots\text{,}\bm{w}_M^T]$ is the entire transmit beamforming vector of $M$ APs. Then, $R_\text{sec}$, $\text{C6}$, $\text{C1}$, and $\text{C3}$ can be respectively reformulated as $\breve{R}_\text{sec}=\log_2(1+\frac{\text{Tr}(\bm{H}_{p}\bm{W})}{\alpha\text{Tr}(\bm{G}_{p}\bm{W}) + \sigma_p^2 })- \log_2(1+\frac{\text{Tr}(\bm{H}_{e}\bm{W})}{\alpha\text{Tr}(\bm{G}_{e}\bm{W}) + \sigma_e^2 })$, $\text{C7:~}\frac{\alpha\text{Tr}(\bm{G}_{s}\bm{W})}{\sigma_s^2}\geq \beta_c^*$, $\text{C8:~} \text{Tr}(\bm{H}_s\bm{W})-(2^{R_\text{th1}}-1)\alpha\text{Tr}(\bm{G}_s\bm{W})\geq(2^{R_\text{th1}}-1)\sigma_s^2$, and $\text{C9:~} \text{Tr}(\bm{W}_m) \le P_\text{{max}}$, where $\bm{H}_\xi=\bm{h}_\xi\bm{h}_\xi^H$, $\bm{G}_\xi=\bm{g}_\xi\bm{g}_\xi^H$, $\bm{W_m} = \bm{w}_m\bm{w}_m^H$ indicates the $m$-th partitioned matrix on the diagonal of $\bm{W}$, $\bm{h}_\xi^H = [\bm{h}_{1\text{,}\xi}^H \text{,}\bm{h}_{2\text{,}\xi}^H\text{,}\cdots\text{,}\bm{h}_{M\text{,}\xi}^H]^H$, and $\bm{g}_\xi = [\bm{H}_{1}\bm{g}_{b\text{,}\xi}\text{;}\bm{H}_{2}\bm{g}_{b\text{,}\xi}\text{;}\cdots\text{;}\bm{H}_{M}\bm{g}_{b\text{,}\xi}]$.
Then, this sub-problem of optimizing transmit beamforming can be stated as
	\begin{equation}\tag{$\textbf{P2}$} 
	\begin{aligned}
		\max_{\bm{W}} ~~& \breve{R}_\text{sec}\\ 
		\text{s.t.}~~ & \text{C7}\text{,}\ \text{C8}\text{,}\  \text{C9}\text{,}\ \text{C10}\text{,}\ \text{C11}.
	\end{aligned}
\end{equation}

To simplify the notation, $\breve{R}_\text{sec}$ is recast as $\breve{R}_\text{sec}=g_1(\bm{W})-g_2(\bm{W})-g_3(\bm{W})+g_4(\bm{W})$, where $g_1(\bm{W}) = \log_2(\alpha\text{Tr}(\bm{G}_p\bm{W})+\sigma_p^2+\text{Tr}(\bm{H}_p\bm{W}))$, $g_2(\bm{W}) = \log_2(\alpha\text{Tr}(\bm{G}_p\bm{W})+\sigma_p^2)$, $g_3(\bm{W}) = \log_2(\alpha\text{Tr}(\bm{G}_e\bm{W})+\sigma_e^2+\text{Tr}(\bm{H}_e\bm{W}))$, and $g_4(\bm{W}) = \log_2(\alpha\text{Tr}(\bm{G}_e\bm{W})+\sigma_e^2)$.
To solve the non-convexity of $\breve{R}_\text{sec}$, {the upper bounds of $g_2(\bm{W})$ and $g_3(\bm{W})$ are obtained by their first-order Taylor approximations at any feasible point $\bm{W}^{(\bar{t})}$ }in the $\bar{t}$-th iteration with the SCA technique \cite{ZHOUC}. Specifically, 
$g_2(\bm{W})\leq  \log_2(\alpha\text{Tr}(\bm{G}_p\bm{W}^{(\bar{t})})+\sigma_p^2)
 +\frac{\alpha\text{Tr}(\bm{G}_p(\bm{W}-\bm{W}^{(\bar{t})}))}{(\alpha\text{Tr}(\bm{G}_p\bm{W}^{(\bar{t})})+\sigma_p^2)\ln2} \triangleq \bar{g}_2(\bm{W})$,   
 $g_3(\bm{W})\leq \log_2(\alpha\text{Tr}(\bm{G}_e\bm{W}^{(\bar{t})})+\sigma_e^2+\text{Tr}(\bm{H}_e\bm{W}^{(\bar{t})}))\notag+\frac{\alpha\text{Tr}(\bm{G}_e(\bm{W}-\bm{W}^{(\bar{t})}))+\text{Tr}(\bm{H}_e(\bm{W}-\bm{W}^{(\bar{t})}))}{(\alpha\text{Tr}(\bm{G}_e\bm{W}^{(\bar{t})})+\sigma_e^2+\text{Tr}(\bm{H}_e\bm{W}^{(\bar{t})}))\ln2} \triangleq \bar{g}_3(\bm{W})$.
Let $\tilde{R}_\text{sec}=g_1(\bm{W})-\bar{g}_2(\bm{W})-\bar{g}_3(\bm{W})+g_4(\bm{W})$. 
 By adopting the SDR technique to relax $\text{C10}$, we can utilize the CVX tool to solve th relaxed version and derive the rank-one solution $\tilde{\bm{W}}$ \cite{LYUB111}. The sub-optimal solution of \textbf{P2}, denoted by $\bm{W}^*$, can finally be derived by iteratively implementing the SCA method.
Performing singular value decomposition on $\bm{W}^*$ allows us to obtain the optimal transmit beamforming vector, denoted by $\bm{w}^*$.


\subsection{Optimization of  the MAs' positions}
\label{OpMA}
In this subsection, we continue to optimize the positions of the MAs when  the transmit beamforming is given. 
The sub-problem {is formulated as}
\begin{equation}\tag{$\textbf{P3}$} 
	\begin{aligned}
		\max_{ \bm{t}} ~~& \tilde{R}_\text{sec} \\ 
		\text{s.t.}~~ & \text{C4} \text{,}\ \text{C5} \text{,}\ \text{C7} \text{,}\ \text{C8}. \\
	\end{aligned}
\end{equation}
Due to the huge solution space provided by $\bm{t}$, directly searching for the optimal solution of \textbf{P3} incurs extremely a high computational complexity. To avoid this issue, the GA-PSO algorithm \cite{GAPSO1} is applied since it can efficiently navigate the complex solution space and yield solutions that balance exploration and exploitation \cite{GAPSO1}.
Algorithm  \ref{AlgorithmA} summarizes the solving process of \textbf{P3} using the GA-PSO algorithm.
In Algorithm \ref{AlgorithmA}, we  firstly introduce $Q$ particles and initialize their positions and velocities to $\mathcal{A}^{(0)} = \{\bm{\hat{\psi}}_1^{(0)}\text{,} \bm{\hat{\psi}}_2^{(0)}\text{,}\ldots\text{,} \bm{\hat{\psi}}_Q^{(0)} \}$ and $\mathcal{V}^{(0)} = \{\bm{v}_1^{(0)}\text{,} \bm{v}_2^{(0)}\text{,}\ldots\text{,} \bm{v}_Q^{(0)} \}$. {Note that each  particle denotes a possible position solution for all MAs, structured as follows: 
$\bm{\hat{\psi}}_q^{(0)} = 
\left[
\begin{array}{c}
	\underbrace{\bm{\hat{\psi}}_{q\text{,}1\text{,}1}^{(0)}\text{;}\cdots\text{;}\bm{\hat{\psi}}_{q\text{,}1\text{,}N}^{(0)}}_{\text{MAs of the }  1 \text{-th AP}}\text{;}  \cdots\text{;}
	\underbrace{\bm{\hat{\psi}}_{q\text{,}M\text{,}1}^{(0)}\text{;}\cdots\text{;}\bm{\hat{\psi}}_{q\text{,}M\text{,}N}^{(0)}}_{\text{MAs of the }M \text{-th AP}} 
\end{array}
\right],$
where $q \in \mathcal{Q}$, and $\mathcal{Q}=\{1\text{,}\cdots\text{,}Q\}$ is the set of  particles. Let  $\bm{\hat{\psi}}_{q\text{,}m\text{,}n}^{(0)} = [x_{q\text{,}m\text{,}n}^{(0)} \text{,}\ y_{q\text{,}m\text{,}n}^{(0)}\text{,}\ z_{q\text{,}m\text{,}n}^{(0)}]^T$ denote the initialization position of the $n$-th MA at the $m$-th AP of the $q$-th particle under the constraint C5.}
In addition, we introduce a fitness function to evaluate the impact of particle positions {in the $s$-the iteration, which is}
 expressed as 
\begin{align}
	\label{FitFun}
	\mathcal{F}(\bm{W}^*\text{,}\ \bm{\hat{\psi}}_q^{(s)}) = \tilde{R}_{\text{sec}}(\bm{W}^*) - \tau \left| \mathcal{R}(\bm{\hat{\psi}}_q^{(s)})\right|.
\end{align}
In \eqref{FitFun}, $\tilde{R}_{\text{sec}}(\bm{W}^*)$ is the maximum secrecy rate derived by the optimal solution of  \textbf{P2}, and $\tau$ is a positive penalty factor ensuring that $\widetilde{R}_{\text{sec}}(\bm{W}^*) - \tau \leq 0$. $\bm{\hat{\psi}}_q^{(s)}$ denotes the updated positions after the $s$-th iteration of applying {the GA-PSO algorithm, where $ s=1\text{,}\cdots\text{,}S$. Let $\mathcal{R}(\bm{\hat{\psi}}_q^{(s)})=\mathcal{R}_1(\bm{\hat{\psi}}_{q\text{,}1}^{(s)})+\mathcal{R}_2(\bm{\hat{\psi}}_{q\text{,}2}^{(s)})+\text{,}\cdots\text{,}+\mathcal{R}_M(\bm{\hat{\psi}}_{q\text{,}M}^{(s)})$, where $\mathcal{R}_m(\bm{\hat{\psi}}_{q\text{,}m}^{(s)})=\{(\bm{\hat{\psi}}_{q\text{,}m\text{,}\hat{\zeta}}^{(s)}\text{,} \bm{\hat{\psi}}_{q\text{,}m\text{,}\hat{\varsigma}}^{(s)}) | \left \| \bm{\hat{\psi}}_{q\text{,}m\text{,}\hat{\zeta}}^{(s)}-\bm{\hat{\psi}}_{q\text{,}m\text{,}\hat{\varsigma}}^{(s)} \right\|_2 < D\text{,}\ \hat{\zeta}\neq \hat{\varsigma}\text{,}\   \hat{\zeta}\in \mathcal{N}\text{,}\ \hat{\varsigma}\in \mathcal{N}\text{,}\ m\in \mathcal{M}\}$ represents the set containing the positions of all pairs of MAs that {defies} constraint C4 at the $m$-th AP. $ \left| \mathcal{R}(\bm{\hat{\psi}}_q^{(s)})\right|$ indicates the cardinality of $\mathcal{R}(\bm{\hat{\psi}}_q^{(s)})$.}


{Subsequently, each particle} updates the velocity and position based on the obtained locally optimal position, i.e., $\bm{\hat{\psi}}_{q\text{,}\text{pbest}}^{(s)}$, and the globally optimal position, i.e., $\bm{\hat{\psi}}_{\text{gbest}}^{(s)}$, which are given by \cite{ZX} 
\begin{align}
	\label{UpdateV}
	\bm{v}_q^{(s+1)} &= \omega \bm{v}_q^{(s)} +c_1 r_1 (\bm{\hat{\psi}}_{q\text{,}\text{pbest}}^{(s)} -\bm{\hat{\psi}}_q^{(s)} ) \notag \\
     & + c_2 r_2 (\bm{\hat{\psi}}_{\text{gbest}}^{(s)} -\bm{\hat{\psi}}_q^{(s)})\text{,} \\
	\label{UpdateT}
	\bm{\hat{\psi}}_q^{(s+1)} &= \bm{\hat{\psi}}_q^{(s)} + \bm{v}_q^{(s+1)}. 
\end{align}		
The inertia weight is denoted by
\begin{align}
	\label{Updatew}
\omega = \omega_{\text{max}}-\frac{(\omega_{\text{max}}-\omega_{\text{min}})s}{S}\text{,}
\end{align}
 which is applied to balance the inertia of particle movement. {$c_1$ and $c_2$ represent the learning factors determing {the step size towards} the optimal position.} $r_1$ and $r_2$ are random parameters generated from $[0\text{,}1]$ uniformly. If the updated particle's position is {out of }the given range, we assign the value of the corresponding range boundary to the current position component \cite{ZX}, i.e.,
 \begin{equation}
	 	  	\label{CheckP}
	 	  	\mu_{m\text{,}n}^{(s)}  = \left \{
	 	  	\begin{aligned}
		 	  	       &\mu_{m\text{,}n}^{\text{min}}\text{,}  ~~~&\text{if}~~ \mu_{m\text{,}n}^{(s)} < \mu_{m\text{,}n}^{\text{min}}\text{,}\\
		             & \mu_{m\text{,}n}^{\text{max}}\text{,} ~~~ &\text{if}~~ \mu_{m\text{,}n}^{(s)} > \mu_{m\text{,}n}^{\text{max}}\text{,}\\
		              & \mu_{m\text{,}n}^{(s)}\text{,}  ~&\text{otherwise}\text{,}
		 	  	\end{aligned}
	 	  	\right.
	 	  \end{equation}
where $\mu=\{x\text{,}y\text{,}z\}$, $ m\in \mathcal{M}$, $ n\in \mathcal{N} $.
 
In step 6 of Algorithm \ref{AlgorithmA}, we update the crossover probability (denoted by $p_{cs}$) and mutation probability (denoted by $p_{mt}$) based on \eqref{ProbUpdate1} and \eqref{ProbUpdate2}, {which are formulated as}
  \begin{align}
	\label{ProbUpdate1}
	p_{cs}^{(s)} = p_{cs}^{\text{max}}-(p_{cs}^{\text{max}}-p_{cs}^{\text{min}}) \frac{s}{S}\text{,}\\
	\label{ProbUpdate2}
	p_{mt}^{(s)} = p_{mt}^{\text{max}}-(p_{mt}^{\text{max}}-p_{mt}^{\text{min}}) \frac{s}{S}\text{,}
\end{align}
where $p_{cs}^{\text{max}}$, $p_{cs}^{\text{min}}$, $p_{mt}^{\text{max}}$, and $p_{mt}^{\text{min}}$ {denote the maximum and minimum crossover probabilities, as well as the maximum and minimum mutation probabilities, respectively.}  
 
From step 10-15 of Algorithm \ref{AlgorithmA}, {we perform crossover and mutation operations, respectively, during the $s$-th iteration.} Specifically, when a random number (denoted by $\eta_1^{(s)} \in [0\text{,} 1]$) is less than the crossover probability, we first perform crossover operation  based on \eqref{Cross1} and \eqref{Cross2}, which are given by
 \begin{align}
 	\label{Cross1}
  {\bm{\hat{\psi}}}_\iota^{(s+1)}=r_3 \bm{\hat{\psi}}_\iota^{(s)}+(1-r_3)\bm{\hat{\psi}}_\varsigma^{(s)}\text{,}\\
  	\label{Cross2}
  {\bm{\hat{\psi}}}_\varsigma^{(s+1)}=(1-r_3) \bm{\hat{\psi}}_\iota^{(s)} +r_3\bm{\hat{\psi}}_\varsigma^{(s)} \text{,}
 \end{align} 
where $\bm{\hat{\psi}}_\iota^{(s)}$ and $\bm{\hat{\psi}}_\varsigma^{(s)}$ represent the $\iota$-th and the $\varsigma$-th particles, which are selected randomly during the $s$-th iteration, $\iota \in \mathcal{Q}$, and $\varsigma \in \mathcal{Q}$. $r_3$ is random parameter generated from $[0\text{,}1]$. After the crossover operation, if the mutation probability is greater than the random number $\eta_2^{(s)} \in [0\text{,} 1]$, 
the single-point Gaussian mutation operation is performed as follows: {
    \begin{align}
   	\label{Mutation}
   	\bm{\hat{\psi}}_{q\text{,}m\text{,}n}^{(s+1)}= \bm{\hat{\psi}}_{q\text{,}m\text{,}n}^{(s)}+\bm{c}_4\text{,}  	
   \end{align}
where $\bm{\hat{\psi}}_{q\text{,}m\text{,}n}^{(s)}$ denotes the position of the $n$-th MA at the $m$-th AP of the $q$-th particle in the $s$-th iteration.} $\bm{c}_4 \in \mathbb{R}^{3\times 1}$, and each of its elements follows  $\mathcal{CN}(0\text{,}1)$.

The crossover and mutation operations enhance the diversity with the population, preventing it from becoming trapped in locally optimal solution, thereby improving the global search capability \cite{GAPSO1}.

  	 \begin{algorithm}
  	\caption{ The GA-PSO algorithm for solving \textbf{P3}.}
  	\label{AlgorithmA}
  	\begin{algorithmic}[1]  

  		\STATE {Initialize  positions $\mathcal{A}^{(s)}$, velocities $\mathcal{V}^{(s)}$, the maximum iteration number $S$, the maximum crossover and mutation probabilities $p_{cs}^{\text{max}}$ and $p_{mt}^{\text{max}}$, the minimum crossover and mutation probabilities $p_{cs}^{\text{min}}$ and $p_{mt}^{\text{min}}$, and set the iteration index $s=0$.}
  		\STATE{Calculate $\mathcal{F} (\bm{W}^*\text{,}\ \bm{\hat{\psi}}_q^{(s)})$ based on \eqref{FitFun}.} 
  		\STATE{Let  $\bm{\hat{\psi}}_{q\text{,}\text{pbest}}^{(s)} = \bm{\hat{\psi}}_q^{(s)}$ and $\bm{\hat{\psi}}_{\text{gbest}}^{(s)} = \arg \max \{ \mathcal{F} (\bm{W}^*\text{,}\ \bm{\hat{\psi}}_1^{(s)})\text{,}\cdots\text{,} \mathcal{F} (\bm{W}^*\text{,}\ \bm{\hat{\psi}}_Q^{(s)}) \}$.}
  	
  		\WHILE{$s \leq S$}
  			\STATE{Update $\omega$ according to \eqref{Updatew}.}
  			\STATE{Update $p_{cs}^{(s)}$ and $p_{mt}^{(s)}$ according to \eqref{ProbUpdate1} and \eqref{ProbUpdate2}, respectively. }
 			\FOR{$q=1:Q$}
 				\STATE {Update $\bm{v}_q^{(s+1)}$  based on \eqref{UpdateV}.} 
 				\STATE {Update $\bm{\hat{\psi}}_q^{(s+1)}$  based on \eqref{UpdateT} and \eqref{CheckP}.}
 				\IF{$p_{cs}^{(s)} \geq  \eta_1^{(s)}$}
 					\STATE{{Perform the crossover} operation according to \eqref{Cross1} and \eqref{Cross2}.}
 				\ENDIF
 				\IF{$p_{mt}^{(s)} \geq  \eta_2^{(s)}$}
 					\STATE{{Perform the mutation} operation according to \eqref{Mutation}.}
 				\ENDIF
 				\STATE{Evaluate the fitness of the $q$-th particle based on \eqref{FitFun}.}
	 			\IF{$\mathcal{F} (\bm{W}^*\text{,}\ \bm{\hat{\psi}}_q^{(s+1)}) > \mathcal{F} (\bm{W}^*\text{,}\ \bm{\hat{\psi}}_{q\text{,}\text{pbest}}^{(s+1)}) $}
	 					\STATE{Update $\bm{\hat{\psi}}_{q\text{,}\text{pbest}}^{(s+1)} = \bm{\hat{\psi}}_q^{(s+1)}$.}
	 			\ENDIF
	 			\IF{$\mathcal{F} (\bm{W}^*\text{,}\ \bm{\hat{\psi}}_q^{(s+1)}) > \mathcal{F} (\bm{W}^*\text{,}\ \bm{\hat{\psi}}_{\text{gbest}}^{(s+1)})$}
						\STATE{Update $\bm{\hat{\psi}}_{\text{gbest}}^{(s+1)} = \bm{\hat{\psi}}_q^{(s+1)}$.}
				\ENDIF 				
  		  	\ENDFOR
  		\ENDWHILE
	\STATE{Return {$\bm{t}={\bm{\hat{\psi}}_{\text{gbest}}^{(S)}}$.}}

  	\end{algorithmic}
  \end{algorithm}

\subsection{The analysis of the proposed AO algorithm}
{In light of the details provided in} Section \ref{OpTB} and Section \ref{OpMA}, we summarize the steps  for solving problem  \textbf{P1} in Algorithm \ref{AlgorithmB} in this section. The computational complexity of Algorithm \ref{AlgorithmB} is $\mathcal{O}( B_1 (B_2M^{4.5}N^{4.5} \log(\frac{1}{\epsilon_2})+ SQ))$ according to \cite{DDZZ}, \cite{LYUB111}, and \cite{ZX}, where $B_1$ and $B_2$ respectively denote the iteration counts for implementing the AO framework and applying the SCA method to solve \textbf{P2}, and $\epsilon_2$ is the computational accuracy required for the SCA method.

  	 \begin{algorithm}
  	\caption{  Algorithm for solving \textbf{P1}.}
  	\label{AlgorithmB}
  	\begin{algorithmic}[1]  
  		\STATE {Initialize the  MAs' positions $\bm{t}^{(0)}$, the convergence threshold $\epsilon$ for the AO framework, the convergence threshold $\epsilon_2$ for the SCA method, and set the AO iteration index $\nu=0$.}
  		\REPEAT
           \STATE{$\nu=\nu+1$.}
           \STATE{Initialize the transmit beamforming matrix $\bm{W}^{(\tilde{\varsigma})}$ and set the SCA iteration index $\tilde{\varsigma}=0$ .}
           \REPEAT
                \STATE{$\tilde{\varsigma}=\tilde{\varsigma}+1$.}
                \STATE{Update $\bm{W}^{(\tilde{\varsigma} +1)}$ by solving \textbf{P2}.}
           \UNTIL{the convergence $\epsilon_2$ is achieved.}
           \STATE{Obtain $\bm{W}^{*}$.}

           \STATE{Update $\bm{t}^{(\nu)}$ by implementing Algorithm \ref{AlgorithmA}.}
  		\UNTIL{ {$R_\text{sec}^{(\nu +1)} -R_\text{sec}^{(\nu)}\le \epsilon$.}}
       \STATE{Return $\bm{W}$ and $\bm{t}$.}
  	\end{algorithmic}
  \end{algorithm}

\section{Numerical Results}\label{section-IV}
{This section conducts numerical simulations to evaluate} the performance of the proposed MA empowered scheme with {the GA-PSO algorithm.} The parameters are configured as follows: $M=\text{3}$, $N=\text{8}$, $L = 4$, $B=100$ kHz, $P_\text{max} = 35$ dBm, $\lambda =0.1$ m, $D=0.5 \lambda = 0.05$ m, $\alpha=0.8$, $\sigma_\xi^2=-40$ dBm, {$R_{\text{th1}}=100$ kbits/s, $R_{\text{th2}}=600$ kbits/s,} $Q=300$, $S=300$, $\tau = 100$, $c_1=c_2=1.4$, $\omega_{\text{max}}=0.9$, $\omega_{\text{min}}=0.4$, $p_{cs}^{\text{max}}=0.95$, $p_{cs}^{\text{min}}=0.2$, $p_{mt}^{\text{max}}=0.1$, $p_{mt}^{\text{max}}=0.01$, and $\epsilon=\epsilon_2=10^{-2}$.
{The BD, the PU,} the SU, and the Eve are located at (-5 m, 5 m, 10 m), (\text{0 m}, \text{10 m}, \text{10 m}), (\text{5 m}, \text{5 m}, \text{10 m}), (\text{1 m}, \text{9 m}, \text{10 m}), respectively. 
 Moreover, the APs are uniformly arranged within a square region measuring 100 meters by 100 meters. This region  is oriented perpendicular to the $z$-axis, with the center located at (\text{0 m}, \text{-50 m}, \text{10 m}). 
 The movable area for MAs at the $m$-AP is oriented perpendicular to the $y$-axis,  and modeled  as a square region with dimension $[-\frac{A}{2}\text{,}\frac{A}{2}]\times [-\frac{A}{2}\text{,}\frac{A}{2}]$, where $A = 6\lambda$. {The amounts of} transmit and receive paths are configured to be equal, i.e., $L_{t\text{,}\kappa}^m = L_{r\text{,}\kappa}^m =L_{t\text{,}\xi}^b= L_{r\text{,}\xi}^b= \bar{L}=10$. Denote {$\text{diag}\{[\widehat{a}_{l\text{,}1}\text{,} \cdots\text{,} \widehat{a}_{l\text{,}\bar{L}}]\}$ and $\text{diag}\{[\widehat{a}_{\xi\text{,}1}\text{,} \cdots\text{,} \widehat{a}_{\xi\text{,}\bar{L}}]\}$
as the path-response matrices $\bm{\Sigma}_l$ and $\bm{\Sigma}_{\xi}$, respectively, where each element satisfies the $\mathcal{CN}(0\text{,}c_0\cdot d^{-\varrho}/\bar{L})$.}  $c_0 = -20 $ dB is the path-loss, 
{the distance between two nodes is characterized by $d$,} and $ \varrho = 1.2 $ represents the path-loss exponent. The AoAs and AoDs are uniformly distributed over the range $[-\frac{\pi}{2}\text{,}\frac{\pi}{2}]$. To compare the system performance, we considered the proposed MA empowered scheme {with the PSO algorithm} \cite{HM}, the FPA empowered scheme with the FPAs at the APs,} and the random transmit beamforming scheme as benchmark schemes.

\begin{figure}[t]
	\centering
	\includegraphics[width=0.34\textwidth]{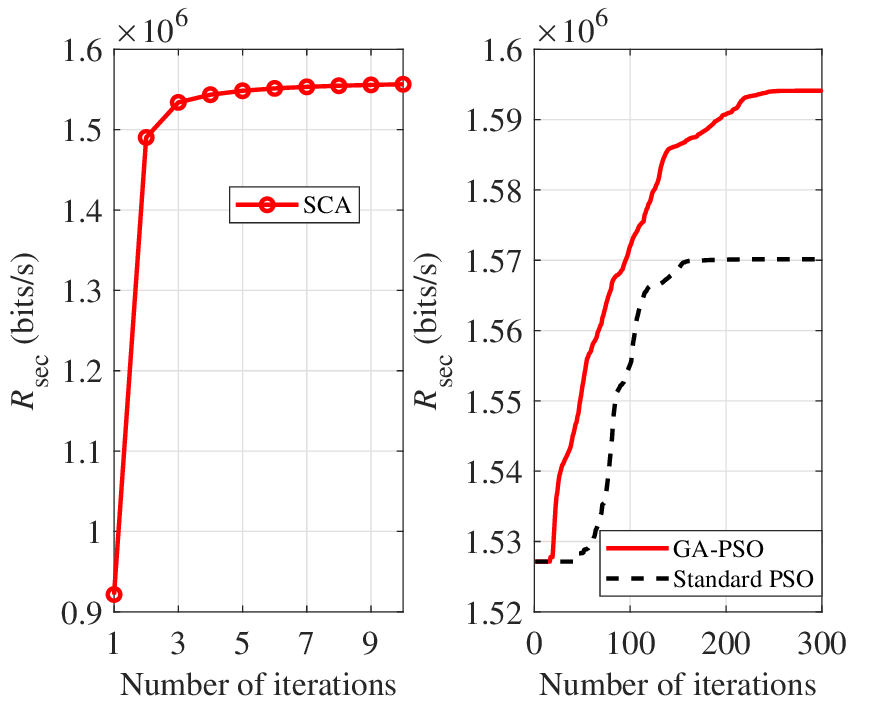} 
	\caption{{Convergence performance} of both the SCA method and the GA-PSO algorithm.}
	\label{fig:SL}
\end{figure}
Fig. \ref{fig:SL} shows {the convergence performance of} both the SCA method and the GA-PSO algorithm. 
It is noted that for the SCA method and the GA-PSO algorithm, the secrecy rate {is a} non-decreasing function with the number of iterations and converges to a determined value within {an acceptable iteration times. As shown in }Fig. \ref{fig:SL}, {the proposed scheme with the GA-PSO algorithm {achieves} a higher secrecy rate, which {showcases that} the GA-PSO algorithm exhibits a superior global search capability in comparison to the PSO algorithm.}

\begin{figure}[t]
  \centering
  \includegraphics[width=0.34\textwidth]{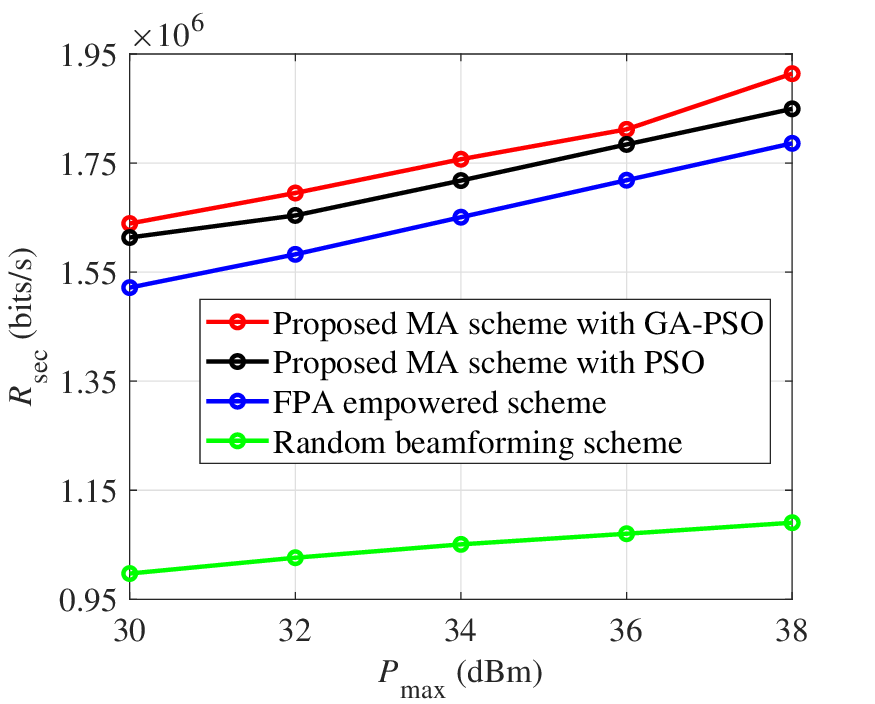}
  \caption{Secrecy rate versus maximum transmit power.}
  \label{fig:power}
\end{figure}

Fig. \ref{fig:power} investigates the secrecy rate versus the each AP's maximum transmit power. Increasing the maximum transmit power straightforwardly improves the secrecy rate across all schemes. In contrast to the FPA empowered scheme, the proposed MA empowered {schemes with the GA-PSO algorithm} and the PSO algorithm result in performance improvements of $7.7\%$ and $6.0\%$, respectively, with $P_\text{max}$ = 30 dBm.
It demonstrates the ability of MAs {to move to particular locations, providing better transmission conditions} for the PU but worsening that for the Eve.

\begin{figure}[t]
	\centering
	\includegraphics[width=0.34\textwidth]{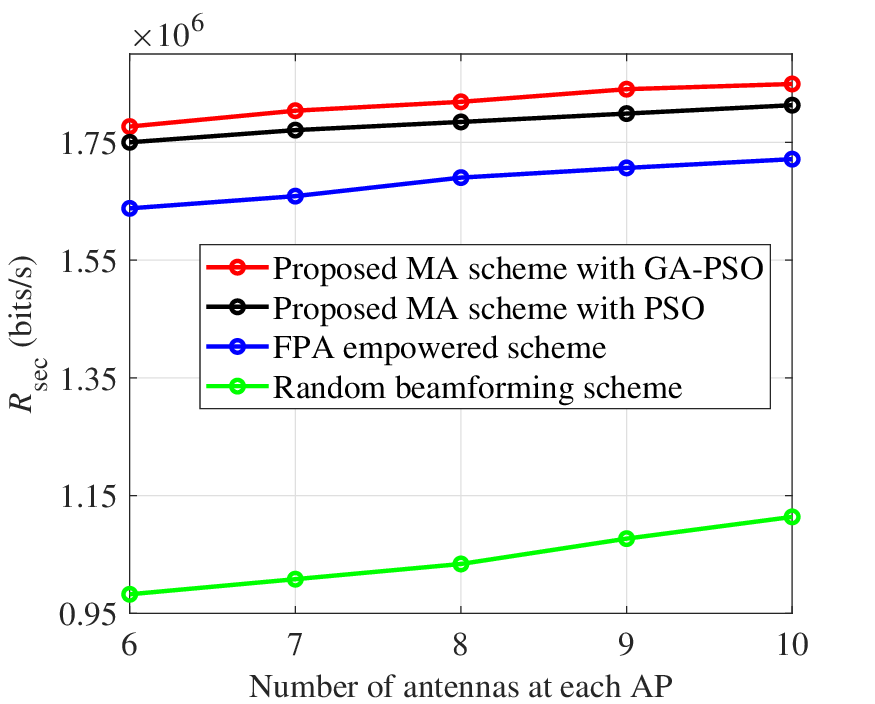}
	\caption{Secrecy rate versus number of antennas at each AP.}
	\label{fig:N}
\end{figure}

Fig. \ref{fig:N} presents {the secrecy rate} versus the number of antennas at each AP, i.e., $N$. Deploying additional antennas at each AP enhances the secrecy rate of all schemes by enabling greater spatial diversity gain.  As shown in Fig. \ref{fig:N},
 the proposed scheme with the GA-PSO algorithm achieves a higher secrecy rate compared to the scheme with PSO algorithm. It is because the involvement of the GA can enhance the population diversity and {strengthen} the global search capabilities of the PSO algorithm.

\begin{figure}[t]
	\centering
	\includegraphics[width=0.34\textwidth]{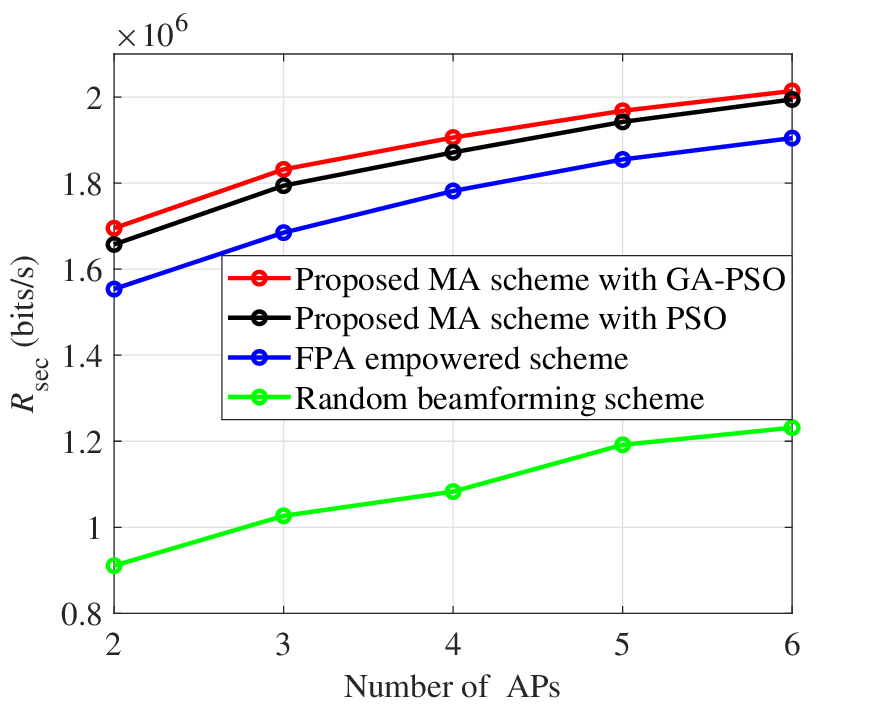}
	\caption{Secrecy rate versus {number} of APs.}
	\label{fig:M}
\end{figure}

Fig. \ref{fig:M} depicts {the secrecy rate} versus the number of the APs. Increasing the number of the APs {positively impacts the secrecy rate.} This enhancement stems from the heightened spatial diversity by introducing multiple independent transmission paths. This diversity effectively mitigates signal attenuation, thereby improving the received signal strength at the PU. {Similar to the illustrations in} Figs. \ref{fig:power} and \ref{fig:N}, our proposed scheme {with the GA-PSO algorithm} exhibits superior performance over the benchmark schemes.
\begin{figure}[t]
	\centering
	\includegraphics[width=0.34\textwidth]{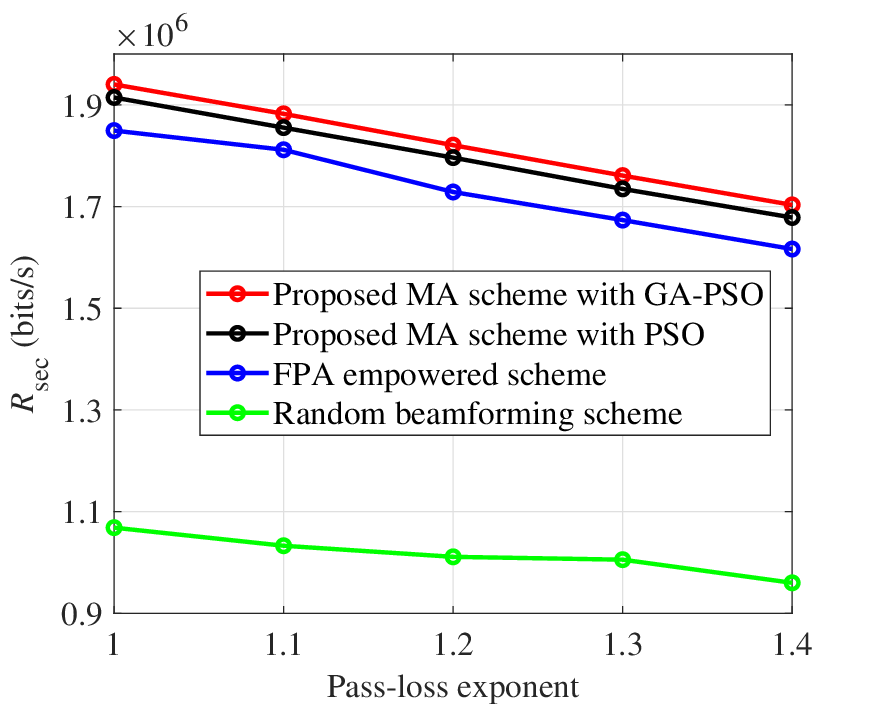}
	\caption{Secrecy rate versus pass-loss exponent.}
	\label{fig:alp}
\end{figure}

The {correlation} between the secrecy rate and the pass-loss exponent is illustrated in Fig. \ref{fig:alp}. As the path loss exponent increases, the secrecy rates of all schemes decline. A higher path loss exponent indicates larger signal attenuation with distance, leading to substantial power loss and constraining long-range transmission capabilities. From Fig. \ref{fig:alp}, it is also evident that the random transmit beamforming scheme performs the worst. {The random beamforming design causes the transmitted signals from the APs to scatter in random directions rather than being directed towards the PU and the BD, thereby reducing the received signal strength at the PU.}


\section{Conclusion}
{In the paper, we have proposed a novel MA empowered {PLS mechanism} for a cell-free SR communication system, where the MAs at the distributed APs are utilized to guarantee the secure transmission from the APs to the PU and enhance the secondary transmission performance. Then, we have studied the secrecy rate maximization problem of the primary transmission for the PU under the constraints on the QoS constraints at the SU. To efficiently address the non-convex nature of the formulated problem,} we have proposed an AO framework based on the SCA method, the SDR technique, and the GA-PSO algorithm to derive an approximately optimal solution.  Finally, numerical results have shown that the proposed MA empowered scheme outperformance the FPA empowered scheme, indicating the potentiality of applying the MAs for performance enhancement. Moreover, compared {to using} the PSO algorithm for the optimization of MA positions, the utilization of the GA-PSO can guarantee an accuracy solution.


		

\begin{thebibliography}{10}
\providecommand{\url}[1]{#1}
\csname url@samestyle\endcsname
\providecommand{\newblock}{\relax}
\providecommand{\bibinfo}[2]{#2}
\providecommand{\BIBentrySTDinterwordspacing}{\spaceskip=0pt\relax}
\providecommand{\BIBentryALTinterwordstretchfactor}{4}
\providecommand{\BIBentryALTinterwordspacing}{\spaceskip=\fontdimen2\font plus
\BIBentryALTinterwordstretchfactor\fontdimen3\font minus
  \fontdimen4\font\relax}
\providecommand{\BIBforeignlanguage}[2]{{%
\expandafter\ifx\csname l@#1\endcsname\relax
\typeout{** WARNING: IEEEtran.bst: No hyphenation pattern has been}%
\typeout{** loaded for the language `#1'. Using the pattern for}%
\typeout{** the default language instead.}%
\else
\language=\csname l@#1\endcsname
\fi
#2}}
\providecommand{\BIBdecl}{\relax}
\BIBdecl

\bibitem{LongIoT}
R.~Long, Y.-C. Liang, H.~Guo, G.~Yang, and R.~Zhang, ``Symbiotic radio: A new
  communication paradigm for passive internet of things,'' \emph{IEEE Internet
  of Things Journal}, vol.~7, no.~2, pp. 1350--1363, 2019.

\bibitem{PLS1}
Z.~Cheng, N.~Li, J.~Zhu, X.~She, C.~Ouyang, and P.~Chen, ``Enabling secure
  wireless communications via movable antennas,'' in \emph{ICASSP 2024-2024
  IEEE International Conference on Acoustics, Speech and Signal Processing
  (ICASSP)}.\hskip 1em plus 0.5em minus 0.4em\relax IEEE, 2024, pp. 9186--9190.

\bibitem{NgoHQ}
H.~Q. Ngo, A.~Ashikhmin, H.~Yang, E.~G. Larsson, and T.~L. Marzetta,
  ``Cell-free massive mimo versus small cells,'' \emph{IEEE Transactions on
  Wireless Communications}, vol.~16, no.~3, pp. 1834--1850, 2017.

\bibitem{ENaye}
E.~Nayebi, A.~Ashikhmin, T.~L. Marzetta, H.~Yang, and B.~D. Rao, ``Precoding
  and power optimization in cell-free massive mimo systems,'' \emph{IEEE
  Transactions on Wireless Communications}, vol.~16, no.~7, pp. 4445--4459,
  2017.

\bibitem{GuoHY}
H.~Guo, Y.-C. Liang, R.~Long, and Q.~Zhang, ``Cooperative ambient backscatter
  system: A symbiotic radio paradigm for passive iot,'' \emph{IEEE Wireless
  Communications Letters}, vol.~8, no.~4, pp. 1191--1194, 2019.

\bibitem{ZChu}
Z.~Chu, W.~Hao, P.~Xiao, M.~Khalily, and R.~Tafazolli, ``Resource allocations
  for symbiotic radio with finite blocklength backscatter link,'' \emph{IEEE
  Internet of Things journal}, vol.~7, no.~9, pp. 8192--8207, 2020.

\bibitem{DDZZ}
Z.~Dai, R.~Li, J.~Xu, Y.~Zeng, and S.~Jin, ``Rate-region characterization and
  channel estimation for cell-free symbiotic radio communications,'' \emph{IEEE
  Transactions on Communications}, vol.~71, no.~2, pp. 674--687, 2022.

\bibitem{PLS}
G.~Hu, Q.~Wu, K.~Xu, J.~Si, and N.~Al-Dhahir, ``Secure wireless communication
  via movable-antenna array,'' \emph{IEEE Signal Processing Letters}, 2024.

\bibitem{TJun}
J.~Tang, C.~Pan, Y.~Zhang, H.~Ren, and K.~Wang, ``Secure mimo communication
  relying on movable antennas,'' \emph{arXiv preprint arXiv:2403.04269}, 2024.

\bibitem{ZHOUC}
B.~Lyu, C.~Zhou, S.~Gong, D.~T. Hoang, and Y.-C. Liang, ``Robust secure
  transmission for active ris enabled symbiotic radio multicast
  communications,'' \emph{IEEE Transactions on Wireless Communications},
  vol.~22, no.~12, pp. 8766--8780, 2023.

\bibitem{ma2023}
W.~Ma, L.~Zhu, and R.~Zhang, ``Mimo capacity characterization for movable
  antenna systems,'' \emph{IEEE Transactions on Wireless Communications}, 2023.

\bibitem{ZHUma}
L.~Zhu, W.~Ma, B.~Ning, and R.~Zhang, ``Movable-antenna enhanced multiuser
  communication via antenna position optimization,'' \emph{IEEE Transactions on
  Wireless Communications}, 2023.

\bibitem{GAPSO1}
F.~Omidinasab and V.~Goodarzimehr, ``A hybrid particle swarm optimization and
  genetic algorithm for truss structures with discrete variables,''
  \emph{Journal of Applied and Computational Mechanics}, vol.~6, no.~3, pp.
  593--604, 2020.

\bibitem{LYUB111}
B.~Lyu, H.~Liu, W.~Hong, S.~Gong, and F.~Tian, ``Primary rate maximization in
  movable antennas empowered symbiotic radio communications,'' \emph{arXiv
  preprint arXiv:2403.14943}, 2024.

\bibitem{GAPSO2}
C.~Roy and D.~K. Das, ``A hybrid genetic algorithm (ga)--particle swarm
  optimization (pso) algorithm for demand side management in smart grid
  considering wind power for cost optimization,'' \emph{S{\=a}dhan{\=a}},
  vol.~46, no.~2, p. 101, 2021.

\bibitem{ZX}
Z.~Xiao, X.~Pi, L.~Zhu, X.-G. Xia, and R.~Zhang, ``Multiuser communications
  with movable-antenna base station: Joint antenna positioning, receive
  combining, and power control,'' \emph{arXiv preprint arXiv:2308.09512}, 2023.

\end{thebibliography}


\begin{thebibliography}{10}
\bibliographystyle{IEEEtran}

\bibitem{YC}
Y. -C. Liang, R. Long, Q. Zhang, and D. Niyato, ``Symbiotic communications: Where marconi meets darwin,'' \emph{ IEEE Wireless Communications}, vol. 29, no. 1, pp. 144-150, Feb. 2022. 

\bibitem{LYC}
Y. -C. Liang, Q. Zhang, E. G. Larsson, and G. Y. Li, ``Symbiotic radio: Cognitive backscattering communications for future wireless networks,'' \emph{ IEEE Transactions on Cognitive Communications and Networking}, vol. 6, no. 4, pp. 1242-1255, Dec. 2020.

\bibitem{LongIoT}
R. Long, Y.-C. Liang, H. Guo, G. Yang, and R. Zhang, ``Symbiotic radio: A new communication paradigm for passive Internet of Things,'' \emph{IEEE Internet of Things Journal}, vol. 7, no. 2, pp. 1350-1363,  Feb. 2020.

\bibitem{HY}
H. Yang, Y. Ye, K. Liang, and X. Chu, ``Energy efficiency maximization for symbiotic radio networks with multiple backscatter devices,'' \emph{ IEEE Open Journal of the Communications Society}, vol. 2, pp. 1431-1444, Jun. 2021. 


\bibitem{DDZZ}
Z. Dai, R. Li, J. Xu, Y. Zeng, and S. Jin, ``Rate-region characterization and channel estimation for cell-free symbiotic radio communications,'' \emph{IEEE Transactions on Communications}, vol. 71, no. 2, pp. 674-687, Feb. 2023. 

\bibitem{FL}
F. Li, Q. Sun, X. Chen, and J. Zhang, ``Spectral efficiency analysis of uplink cell-free massive MIMO symbiotic radio,''  \emph{ IEEE Internet of Things Journal}, vol. 11, no. 2, pp. 3614-3627, Jan. 2024.




\bibitem{PLS1}
Z. Cheng, N. Li, J. Zhu, X. She, C. Ouyang, and P. Chen, ``Enabling secure wireless communications via movable antennas,'' \emph{2024 IEEE International Conference on Acoustics, Speech and Signal Processing (ICASSP)}, pp. 9186-9190, 2024.

\bibitem{ST}
S. Timilsina, D. Kudathanthirige, and G. Amarasuriya, ``Physical layer security in cell-free massive MIMO,'' \emph{2018 IEEE Global Communications Conference (GLOBECOM), Abu Dhabi, United Arab Emirates},  pp. 1-7, 2018.

\bibitem{AAI}
A. Al-Nahari, R. Jäntti, G. Zheng, D. Mishra, and M. Nie, ``Ergodic secrecy rate analysis and optimal power allocation for symbiotic radio networks,'' \emph{IEEE Access}, vol. 11, pp. 82327-82337, Aug. 2023.

\bibitem{ma2023}
W. Ma, L. Zhu, and R. Zhang, ``MIMO capacity characterization for movable antenna systems,"  \emph{ IEEE Transactions on Wireless Communications}, vol. 23, no. 4, pp. 3392-3407, Apr. 2024. 



\bibitem{QQQ}
P. Ramezani, B. Lyu, and A. Jamalipour, ``Toward RIS-enhanced integrated terrestrial/non-terrestrial connectivity in 6G,'' \emph{IEEE Network}, vol. 37, no. 3, pp. 178-185, May/June 2023.



\bibitem{LZHUMA}
L. Zhu, W. Ma, and R. Zhang, ``Modeling and performance analysis
for movable antenna enabled wireless communications,'' \emph {IEEE Transactions on Wireless Communications}, vol. 23, no. 6, pp. 6234-6250, Jun. 2024.

\bibitem{PLS}
G. Hu, Q. Wu, K. Xu, J. Si, and N. Al-Dhahir, ``Secure wireless communication via movable-antenna array,'' \emph{ IEEE Signal Processing Letters}, vol. 31, pp. 516-520, Jan. 2024. 

\bibitem{LYUB111}
B. Lyu, H. Liu, W. Hong, S. Gong, and F. Tian, ``Primary rate maximization in movable antennas empowered symbiotic radio communications,'' \emph{2024 IEEE 99th Vehicular Technology Conference (VTC2024-Spring)},  pp. 1–6,  2024.

\bibitem{ZC}
C. Zhou, B. Lyu, C. You, and Z. Liu, ``Movable Antenna Enabled Symbiotic Radio Systems: An Opportunity for Mutualism,'' \emph{IEEE Wireless Communications Letters}, doi: 10.1109/LWC.2024.3443460, 2024.



\bibitem{ZHOUC}
B. Lyu, C. Zhou, S. Gong, D. T. Hoang, and Y. -C. Liang, ``Robust secure transmission for active RIS enabled symbiotic radio multicast communications,'' \emph{ IEEE Transactions on Wireless Communications}, vol. 22, no. 12, pp. 8766-8780, Dec. 2023. 



\bibitem{ZHUma}
L. Zhu, W. Ma, B. Ning, and R. Zhang, ``Movable-antenna enhanced multiuser communication via antenna position optimization,'' \emph{IEEE Transactions on Wireless Communications}, vol. 23, no. 7, pp. 7214-7229, Jul. 2024.

\bibitem{GAPSO1}
F. Omidinasab and V. Goodarzi, ``A hybrid particle swarm optimization and genetic algorithm for truss structures with discrete variables,'' \emph{Journal of Applied and Computational Mechanics}, vol. 6, no. 3, pp. 593-604, Jul. 2020.

\bibitem{ZX}
Z. Xiao, X. Pi, L. Zhu, X. Xia, and R. Zhang, ``Multiuser communications with movable-antenna base station: Joint antenna positioning, receive combining, and power control,'' [Online]. Available: https://arxiv.org/abs/2308.09512, 2023.

\bibitem{HM}
H. Mao, Y. Liu, Z. Xiao, Z. Han, and X.-G. Xia, ``Joint resource
allocation and 3D deployment for multi-UAV covert communications,” \emph{IEEE Internet of Things Journal}, vol. 11, no. 1, pp. 559-572, Jan. 2024.


%
%
%
%







\end{thebibliography}
\end{document}